# Statistical analysis of the temporal single-photon response of superconducting nanowire single photon detection*

He Yu-Hao(何宇昊), Lv Chao-Lin(吕超林), Zhang Wei-Jun(张伟君), Zhang Lu(张露), Wu Jun-Jie(巫君杰), Chen Si-Jing(陈思井), You Li-Xing(尤立星)†, Wang Zhen(王镇)

State Key Laboratory of Functional Materials for Informatics, Shanghai Institute of Microsystem and Information Technology, Chinese Academy of Sciences, Shanghai 200050, China

**Abstract**

Counting rate is a key parameter of superconducting nanowire single photon detectors (SNSPD) and is determined by the current recovery time of an SNSPD after a detection event. We propose a new method to study the transient detection efficiency (*DE*) and pulse amplitude during the current recovery process by statistically analyzing the single photon response of an SNSPD under photon illumination with a high repetition rate. The transient *DE* results match well with the DEs deduced from the static current dependence of *DE* combined with the waveform of a single-photon detection event. This proves that the static measurement results can be used to analyze the transient current recovery process after a detection event. The results are relevant for understanding the current recovery process of SNSPDs after a detection event and for determining the counting rate of SNSPDs.

**Keywords:** single photon detector, recovery process, kinetic inductance

## 1. Introduction

The superconducting nanowire single-photon detector (SNSPD) is a promising

*Project supported by Strategic Priority Research Program (B) of the Chinese Academy of Sciences (XDB04010200) and National Basic Research Program of China (2011CBA00202) and the National Natural Science Foundation of China (61401441).

†Corresponding author. E-mail: lxyou@mail.sim.ac.cn

technology for near-infrared single photon detection owing to its high system detection efficiency (DE), low dark count rate (DCR), low timing jitter, and high counting rate[1-5]. Many SNSPD-based applications have been demonstrated, such as quantum key distribution (QKD)[6-9], deep space laser communications[10], and time-of-fight laser ranging and imaging[11].

Counting rate is an important parameter of SNSPDs for many of the applications mentioned above. Fundamentally, the SNSPD counting rate is determined by the thermal relaxation time of the hotspot generated after a photon is absorbed by nanowire, which is related to the thermal relaxation process from the nanowire to the substrate. The thermal relaxation time is typically of the order of a few tens of picoseconds[12]. A practical SNSPD has a meander nanowire structure covering a sensitive area of $10 \times 10$ μm$^2$, which results in a large kinetic inductance, $L_k$, of the order of 100 nH[13]. Consequently, the current recovery time of a SNSPD for a photon detection event ($\tau_r = L_k / 50\ \Omega$) is several nanoseconds to several tens of nanoseconds, which is 2–3 orders of magnitude higher than the thermal relaxation time. As a result, the counting rate of a practical SNSPD is determined by $L_k$ rather than by the thermal relaxation time.

During the current recovery process, the SNSPD is not completely "dead" and may still be able to partially detect photons. The detection property during the current recovery process provides information on how the current recovers in the SNSPD. Moreover, it may provide interesting insights on the evaluation of the counting rate and on how to improve the counting rate.

Kerman et al.[13] conducted an experiment to illuminate devices with optical pulse pairs, and then they measured the probability that both pulses were detected as a function of the pulse separation. Zhao et al.[14] reported the DE measurement data during the current recovery process using the horizontal histogram of the pulses that occurred after the triggered pulse. In this paper, we introduce a new method based on the statistical analysis of a long-term photon response waveform, collected by a sophisticated oscilloscope (Tek DSA 71254) during illumination of the SNSPD by a laser with a high repetition rate. With this method, we are able to study both DE and the amplitude of the photon response during the current recovery process; i.e., the transient property of an SNSPD after a single-photon detection event has occurred. The results match well with the measured current-dependent system DE and can be quantitatively explained by a simplified model based on current recovery as determined by the kinetic inductance.

## 2. SNSPD system and the Device Characterization

NbN film with a thickness of 7 nm was magnetron sputtered on a double-side oxidized Si wafer. A meander nanowire, covering an active area of 15 × 15 μm$^2$, was fabricated using electron beam lithography and reactive ion etching. The width and pitch of the nanowire were 90 and 240 nm, respectively, which gave a filling factor of 37.5%. The superconducting critical temperature $T_c$ of the device was 7.2 K, and the room-temperature resistance was 5.15 MΩ. We used a Gifford–McMahon (GM) cryocooler to cool the SNSPD down to 2.21 ± 0.02 K, and the switching current ($I_{sw}$) of the SNSPD was 20.3 μA. The kinetic inductance was 1200 nH, obtained by measuring the phase of the reflection coefficient $S_{11}$ as a function of frequency, as determined by a network analyzer. [15]

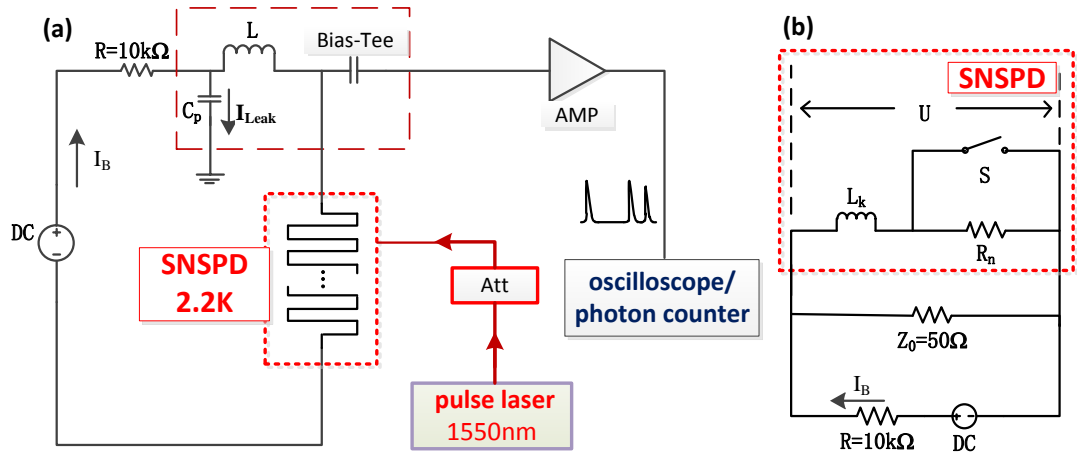

Fig. 1. (a) The SNSPD measurement schematic using a quasi-constant-current bias. The red lines represent the light route and the black lines represent the electricity route. (b) The equivalent SNSPD circuit schematic with a quasi-constant-current bias.

Fig. 1 (a) shows a schematic for measuring SNSPD. A quasi-constant-current bias was applied to the SNSPD using an isolated voltage source in series with a 10 kΩ resistor. A 1550-nm-wavelength picosecond pulsed laser with a variable repetition rate was used as the photon source. The SNSPD device was illuminated from the backside of the Si substrate through a single-mode lensed fiber, which ensured a high optical coupling efficiency. [16] The photon-response signal of the SNSPD could be directed to either the photon counter or oscilloscope. Figure 2(a) shows the current dependence of the system DE and DCR. The system DE reached 16% and 45% at 10 and 100 Hz, respectively.

To study the current relaxation of the SNSPD, the output single-photon response waveform was recorded and is shown in Figure 2(b) (blue line). The pulse amplitude is represented as $V_{\text{pulse}} = (I_B - I_r) \times Z_0 \times G$, where $I_B$ is the bias current; $I_r$ is the return current; $Z_0$ is the rf impedance of the coaxial cable (50 Ω); and $G = 260$ is the measured gain of the signal path[13]. We could also deduce the current through the device as $I = I_B - U / Z_0$ (the blue line shown in Figure 2(c)), where $U$ is the transient voltage of the SNSPD, which is equal to the recorded transient voltage of the single-photon response waveform divided by the gain. To further investigate the electrical mechanism of the SNSPD, the circuit model of the SNSPD is shown in Fig. 1 (b). The switch represents the SNSPD switching between the superconducting and resistive state. The variable $R_n$ denotes the resistance of the resistive domain during single-photon detection. Because the resistive domain appears and disappears on the thermal relaxation time scale, we regard $R_n$ as a time-invariable value (about 450 Ω)[17] to simplify the calculation. By defining the moment that the device absorbs the photon as the start time, the current through the device could be derived from the circuit model and is presented as follows:

$$\begin{cases} I_{\text{fall}} = I_B - (I_B - I_r) \times e^{-\frac{Z_0 + R_n}{L_k} \times t} & (0 \leq t < t_{\text{peak}}) \\ I_{\text{rise}} = I_r + (I_B - I_r) \times e^{-\frac{Z_0}{L_k}(t - t_{\text{peak}})} & (t \geq t_{\text{peak}}) \end{cases} \quad (1)$$

where $I_{\text{fall}}$ and $I_{\text{rise}}$ represent the current through the SNSPD device during the fast falling and slow rising edge, respectively; $t_{\text{peak}}$ is the time taken for the device current to reach its minimum value (~4 ns), as shown in Fig. 2(c). By solving the equations (1) above, we calculated the single-photon response waveform and the current through the SNSPD shown as red dashed lines in Fig. 2(b) and (c). There are some discrepancies between the experimental waveform and calculated ones, which may be explained by the simplified model adopted. For example, a small capacitor $C_p$ (190 nF) exists in the bias-tee, which was neglected in the calculation, and the leaking current $I_{Leak}$ through $C_p$ may have caused a smaller voltage drop during the falling edge of the waveform. Additionally, the non-ideal bandpass characteristics of the wideband amplifier (RF Bays LNA650: (30 k - 600 MHz), $G$ = 50 dB, $NF$ = 2.2 dB) may have also contributed to the discrepancies.

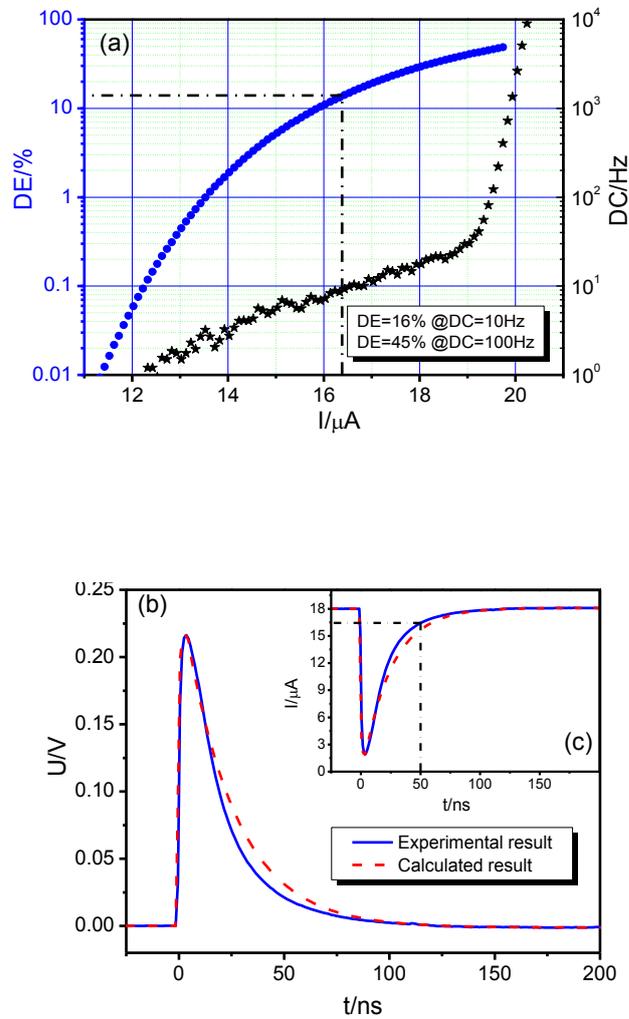

Fig. 2. (a) The system DE (blue dots) and DCR (black stars) of the SNSPD versus the bias current *I*. (b) The experimental single-photon detection waveform (blue solid line) and the calculated single-photon detection waveform (red dashed line). The experimental waveform was obtained by averaging over 10000 times to minimize the noise influence. (c) The current flowed through the SNSPD, based on experimental and calculated waveforms in (b).

## 3. Experiment and Results

The advanced digital phosphor oscilloscope (Tek DSA71254) provides us several functions to study the transient single photon response. For example, the fast acquisition mode allows us to superimpose multiple single-photon detection waveforms directly upon a signal triggered by the rising edge of a waveform. A screenshot of the fast acquisition result is shown in Fig. 3. The bias current was 18.0 μA (0.887 $I_{sw}$), and the system DE was measured to be 29.1%. The repetition rate of the laser was set as 1 GHz, and the intensity was strongly attenuated to 10 M photons per second. We observed a "dead" time zone of ~25 ns without any subsequent photon detection events from Fig. 3. Thereafter, a few detection events started to appear, and the event quantity increased with the elapse of time. After ~100 ns, the quantity of the detection events seemed to saturate. This result indicates that there is a "dead" time zone in the beginning of the current recovery process, after which the single-photon detection ability recovers continuously, and finally the DE attains a normal value. Therefore, DE may decrease when SNSPD is measured with either the high-repetition-rate pulsed laser or the high-intensity continuous-wave laser. Another noticeable result is that the amplitude of the first few pulses seems to be smaller than the normal value, which has not been previously reported.

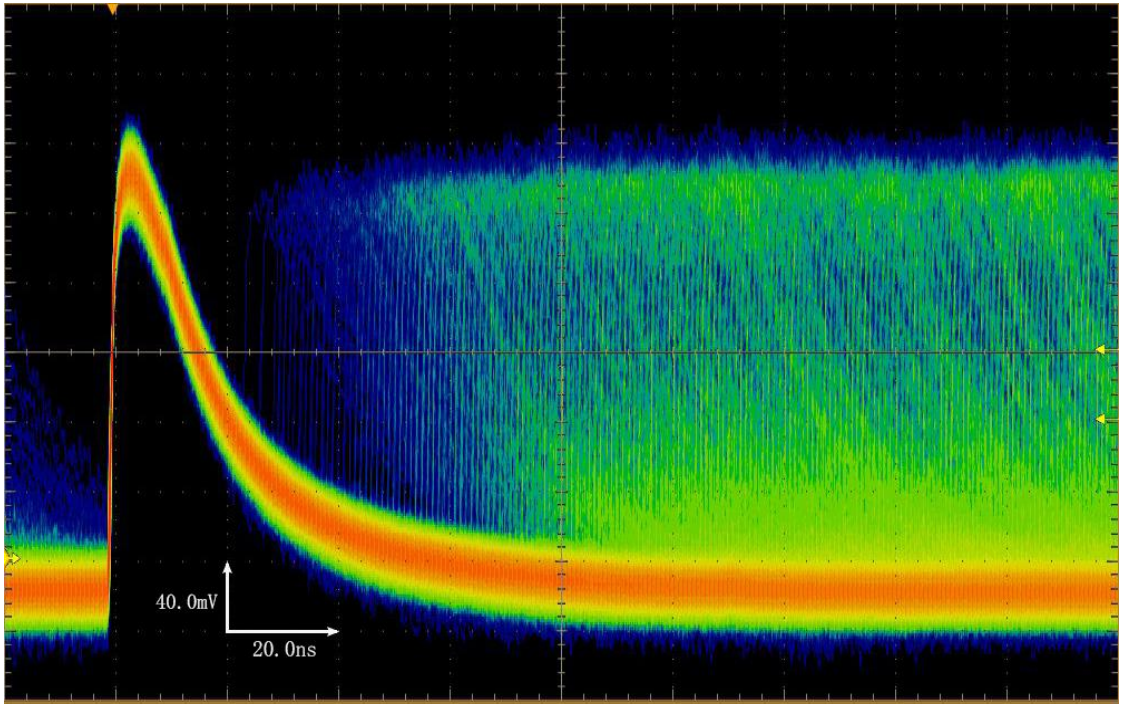

**Fig. 3. A screenshot of the continuous single-photon response waveforms. The red, green, and blue colors indicated the events occured with high, medium, and low probabilities, respectively.**

To quantitatively analyze the detection ability during the current recovery process of the SNSPD, we used the oscilloscope to record long-term waveform data and performed the statistical analysis offline. The method is briefly described as follows. First, data over 4 seconds was recorded, which included $10^5$ periods. Each period was 40 μs and over 100 detection events were recorded in each period. The data were smoothed to reduce noise. The time corresponding to the maximum derivative of the data was regarded as the arrival time of each detection event. The maximum voltage in 5 ns after the arrival time is chosen as the amplitude of each detection event, which is correct since the duration of the rising edge is typically ~ 3 ns. We collected the arrival time and the amplitude of each detection event for 200 ns after a random detection event, and regarded this as one record. The DE and amplitude of the detection event during the current recovery process of the SNSPD were determined by the statistical analysis of $N$ (~4 $\times 10^5$) records. The *DE* was equal to the sum of the events in each time bin (1 ns), divided by the product of N and the average photon per pulse. The histogram of the *DE* for different arrival times after a single-photon detection event is shown in Fig. 4(a).

In fact there is another possible method which may produce time-dependent DE of SNSPD. The blue line in Fig. 2(c) gives us the current through SNSPD during the relaxation process. Fig. 2(a) shows the current dependence of DE. If we combine the data from Fig. 2(a) and (c), we may obtain the time dependence of the *DE*, which is shown as the red line in Fig. 4(a). For example, when $t = 50$ ns, we obtained $I = 16.4$ μA ($0.808\ I_{sw}$) from Fig. 2(c), and then we obtained $DE = 13.9\%$ from Fig. 2(a). These results matched well with the statistically analyzed results in Fig. 4(a). This indicates that the detection properties during the current recovery period can be obtained directly by the static measurement of the current dependence of *DE* combined with the waveform of a single-photon detection event. The slight discrepancy may be explained by the differences between the current obtained from the output signal pulse (shown as the blue curve in Fig. 2(c)) and the real current through the nanowire. Because the current shown in Fig. 2 (c) was obtained without considering the influence of the leaking current in the bias tee as well as the bandwidth limitation of the amplifier.

  The amplitude of the detection event that occurred during the current recovery period was obtained by averaging the amplitude of all events in the same time bin (Fig. 4(b)). The values of the voltage pulse amplitudes in each time bin have a Gaussian distribution. The error bar in the figure is given by the standard deviation. As compared with the first pulse, a smaller and gradually increasing amplitude was observed, which was consistent with the observation in Fig. 3. The origin of the abnormal fluctuation during 20 - 25 ns was mainly caused by the limited number of collected data because of a very low DE (less than 0.2 %). However, the circuit simulation in Fig. 2(b) yielded the same amplitude for all detection pulses. Some components in the circuit may cause this discrepancy. For example, $I_{Leak}$ and/or the decay of the first voltage pulse in the recovery process may cause the amplitude of the second pulse to be lower than that of the first if the two pulses are very close. Moreover, the amplifier's output power and bandwidth limitations may have also contributed to the smaller amplitude of the second pulse.

  Fig. 4(a) gives us an intuitive way to evaluate the counting rate of the SNSPD. For example, we can define the counting rate when the *DE* decreases to 50% of its maximum value, which is 14.5%. The corresponding time delay is 51 ns, which corresponds to a counting rate of ~20 MHz. As the results of the statistical analysis match well with the calculated results from Fig. 2(a) and (c), an easy way to determine the counting rate is to analyze the red curve in Fig. 4(a), which was obtained from the traditional curves of *DE* versus $I_B$ and the output pulse waveform.

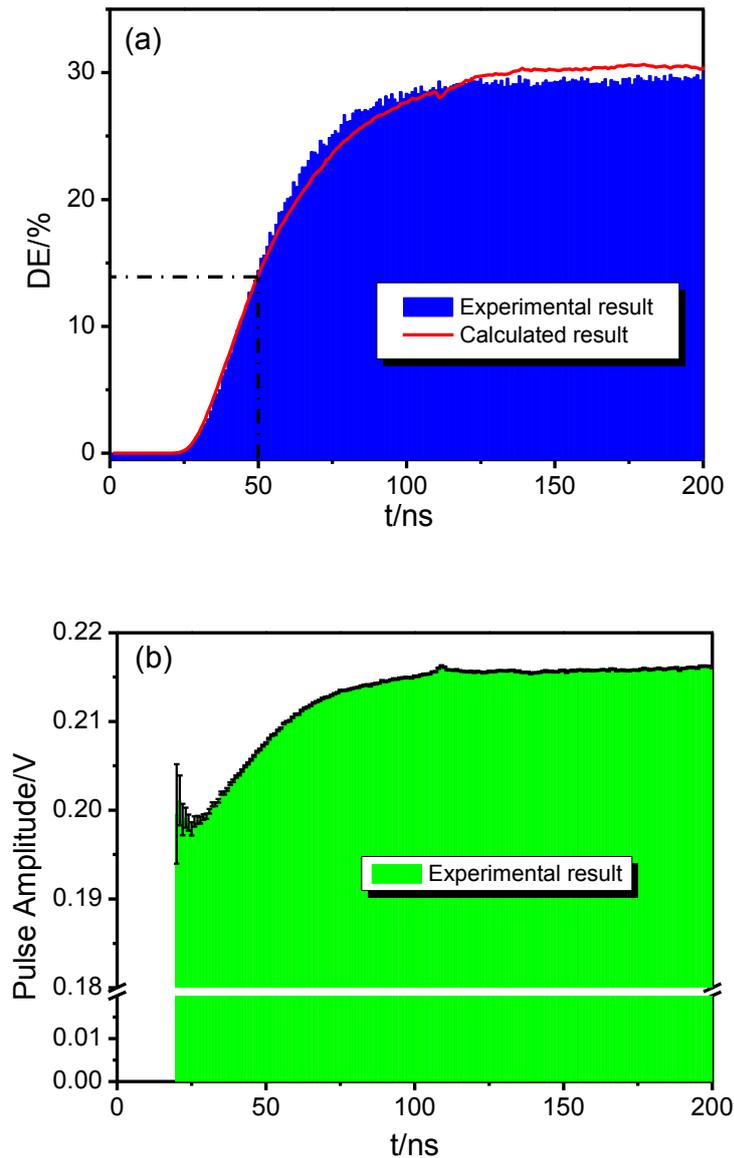

**Fig. 4. (a) The time dependence of the *DE* obtained from the statistical analysis (blue histogram) and the calculated result based on measured data in Fig. 2 (blue curve). (b) The time dependence of the normalized amplitude of the detection event obtained from the statistical analysis (green histogram).**

## 4. Conclusion

In summary, by statistically analyzing the single-photon detection events collected with an oscilloscope, we extensively studied the transient detection property of SNSPDs after a single-photon detection event occurred. A dead time zone, followed by a suppressed *DE* zone, was observed during the current recovery process. The *DE* results matched quantitatively well with the static measurement results of the current dependence of *DE* and the pulse waveform. These results provide us with a simple method to evaluate the counting rate of the SNSPD.